\documentclass[conference]{IEEEtran}
\usepackage{flushend}

\ifCLASSINFOpdf
  \usepackage[pdftex]{graphicx}
  \DeclareGraphicsExtensions{.pdf,.jpeg,.jpg,.png}
\else
\fi
\usepackage{amsmath}
\ifCLASSOPTIONcompsoc
  \usepackage[caption=false,font=normalsize,labelfont=sf,textfont=sf]{subfig}
\else
  \usepackage[caption=false,font=footnotesize]{subfig}
\fi
\begin{document}
%
% paper title
% Titles are generally capitalized except for words such as a, an, and, as,
% at, but, by, for, in, nor, of, on, or, the, to and up, which are usually
% not capitalized unless they are the first or last word of the title.
% Linebreaks \\ can be used within to get better formatting as desired.
% Do not put math or special symbols in the title.
\title{FemtoDAQ: A Low-Cost Digitizer for SiPM-Based \\ Detector Studies
and its Application \\ to the HAWC Detector Upgrade}

% author names and affiliations
% use a multiple column layout for up to three different
% affiliations
\author{\IEEEauthorblockN{Wojtek Skulski}
\IEEEauthorblockA{SkuTek Instrumentation\\
West Henrietta, New York 14586 \\
Telephone: (585) 444-7074 \\
Email: wojtek@skutek.com}
\and
\IEEEauthorblockN{Andreas Ruben}
\IEEEauthorblockA{W-IE-NE-R Plein and Baus Corp.\\
Springfield, Ohio 45505\\
Telephone: (937) 324-2420\\
Email: aruben@wiener-us.com}
\and
\IEEEauthorblockN{Segev BenZvi}
\IEEEauthorblockA{Department of Physics and Astronomy\\
University of Rochester\\
Rochester, New York 14627\\
Telephone: (585) 276-7172\\
Email: sybenzvi@pas.rochester.edu}}

% conference papers do not typically use \thanks and this command
% is locked out in conference mode. If really needed, such as for
% the acknowledgment of grants, issue a \IEEEoverridecommandlockouts
% after \documentclass

% for over three affiliations, or if they all won't fit within the width
% of the page, use this alternative format:
% 
%\author{\IEEEauthorblockN{Michael Shell\IEEEauthorrefmark{1},
%Homer Simpson\IEEEauthorrefmark{2},
%James Kirk\IEEEauthorrefmark{3}, 
%Montgomery Scott\IEEEauthorrefmark{3} and
%Eldon Tyrell\IEEEauthorrefmark{4}}
%\IEEEauthorblockA{\IEEEauthorrefmark{1}School of Electrical and Computer Engineering\\
%Georgia Institute of Technology,
%Atlanta, Georgia 30332--0250\\ Email: see http://www.michaelshell.org/contact.html}
%\IEEEauthorblockA{\IEEEauthorrefmark{2}Twentieth Century Fox, Springfield, USA\\
%Email: homer@thesimpsons.com}
%\IEEEauthorblockA{\IEEEauthorrefmark{3}Starfleet Academy, San Francisco, California 96678-2391\\
%Telephone: (800) 555--1212, Fax: (888) 555--1212}
%\IEEEauthorblockA{\IEEEauthorrefmark{4}Tyrell Inc., 123 Replicant Street, Los Angeles, California 90210--4321}}

% use for special paper notices
%\IEEEspecialpapernotice{(Invited Paper)}

% make the title area
\maketitle

% As a general rule, do not put math, special symbols or citations
% in the abstract
\begin{abstract}

The FemtoDAQ is a low-cost two channel data acquisition system which we have
used to investigate the signal characteristics of silicon photomultipliers
(SiPMs) coupled to fast scintilators. The FemtoDAQ system can also be used to
instrument low cost moderate performance passive detectors, and is suitable for
use in harsh environments (e.g., high altitude). The FemtoDAQ is being used as
a SiPM test bench for the High Altitude Water Cherenkov (HAWC) Observatory, a
TeV gamma ray detector located 4100 m above sea level. Planned upgrades to the
HAWC array can benefit greatly from SiPMs, a robust, low-voltage, low-cost
alternative to traditional vacuum photomultipliers. The FemtoDAQ is used to
power the SiPM detector front end, bias the SiPM, and digitize the photosensor
output in a single compact unit.

\end{abstract}

% no keywords

% For peer review papers, you can put extra information on the cover
% page as needed:
% \ifCLASSOPTIONpeerreview
% \begin{center} \bfseries EDICS Category: 3-BBND \end{center}
% \fi
%
% For peerreview papers, this IEEEtran command inserts a page break and
% creates the second title. It will be ignored for other modes.
\IEEEpeerreviewmaketitle

\section{Introduction}

The capture, digitization, and postprocessing of waveforms from photosensors is
among the most common measurements performed in nuclear and high-energy
physics. Traditionally such measurements require a significant investment in
space and equipment, with researchers custom designing test benches out of
components found in the laboratory: benchtop or rack-mounted power supplies,
amplifiers, digitizers, readout hardware and software, etc. While this kind of
equipment is accessible to scientists and engineers in most professional
settings, its cost (and the expertise needed to run it) is often beyond the
capabilities of students and educators.

However, the continuing drop in the cost of computers and electronics now makes
it possible to design powerful and compact data acquisition systems using
commercial off-the-shelf components.  In this paper we describe such a system,
called the FemtoDAQ \cite{FemtoDAQ.com}.

The FemtoDAQ is a two-channel data acquisition computer with signal processing
capabilities. It is designed to power and read out silicon photomultipliers
(SiPMs) but can interface with traditional photomultipliers (PMTs) and be used
in a wide variety of experiments. Its compact size makes it easy to transport
and well-suited for remote applications. In addition, a simple Python-based
programming layer makes the unit easy to use in the classroom, but it is
powerful enough to act as a drop-in replacement for a full-fledged laboratory
test bench. We are currently using the FemtoDAQ to characterize silicon
photomultipliers (SiPMs) for the High-Altitude Water Cherenkov (HAWC)
Observatory, a US-Mexican facility built to observe astrophysical gamma rays
\cite{Abeysekara:2013}.

This paper is organized as follows. In Section~\ref{sec:femtodaq} we describe
the components of the FemtoDAQ, the basics of its operation, and a simple
example of its capabilities. In Section~\ref{sec:hawc} we describe the HAWC
detector and a study of SiPMs for use in HAWC. We summarize the status of our
current work and discuss future directions and applications in
Section~\ref{sec:conclusion}.

\section{Description of the FemtoDAQ LV-2}\label{sec:femtodaq}

\begin{figure}[!t]
\centering
\frame{\includegraphics[width=2.5in]{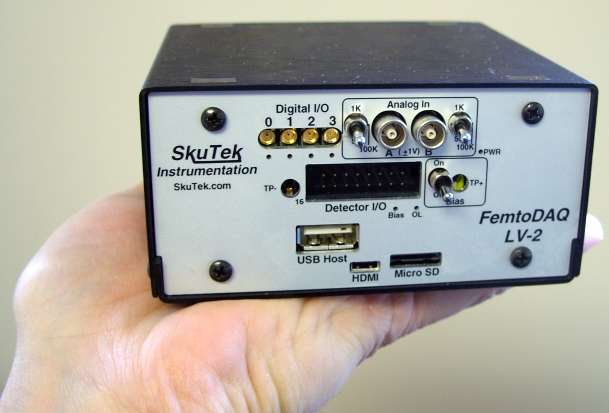}}
\frame{\includegraphics[width=2.5in]{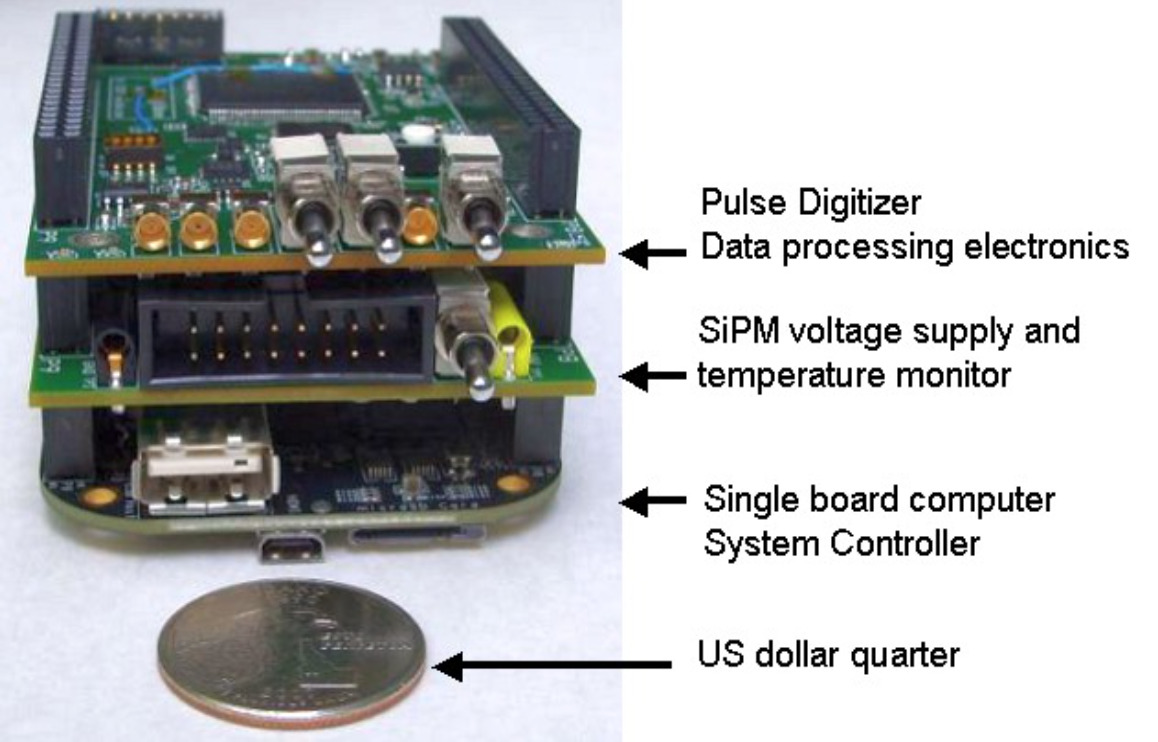}}
\caption{The FemtoDAQ unit (top) and interior boards (bottom).}
\label{fig:box}
\end{figure}

The FemtoDAQ, shown in Fig.~\ref{fig:box}, is a commercial single-board
BeagleBone Black computer \cite{BeagleBone:2016} with two custom-designed
mezzanine boards. The first mezzanine board is specifically designed to
interface with silicon photomultipliers.  SiPMs are being rapidly adopted in
fields such as medical imaging, astronomy, and high-energy physics due to their
low cost, low power and voltage requirements, high gain, and relative
durability compared to PMTs.

The power board in the FemtoDAQ provides 10~V to 90~V DC bias voltage to the
front-end electronics required to operate two SiPMs.  This range makes the
board compatible with commercially available SiPMs from manufacturers such as
Hamamatsu \cite{Hamamatsu:2016} and SensL \cite{SensL:2016}. A ribbon cable
connector is available for interfacing with external devices over SPI and I2C
busses. The busses enable reading of remote temperature sensors to compensate
the SiPM bias for temperature fluctuations and maintain a constant gain.

The second mezzanine board digitizes two analog inputs with 14-bit ADCs
digitizing at 100 MSPS.  An antialiasing Bessel filter is applied to the
waveforms to suppress high-frequency noise and stretch very fast signals.  An
onboard field programmable gate array (FPGA) then performs pulse detection and
extracts pulse characteristics using digital signal processing.  The input
section has a $\pm1$~V range with selectable input impedance: 50~$\Omega$,
1~k$\Omega$, and 100~k$\Omega$. Four logic input/output signals (3.3~V CMOS)
are provided for external triggering or external veto.

While the FemtoDAQ power/bias board has been designed with the testing of SiPMs
in mind, the input board is agnostic to the type of input sensor. The only
limitations are the input impedance levels and signal range. Hence, it can be
used to test SiPMs and similar detectors such as PIN diodes and multi-pixel
photon counters, as well as PMTs (with an appropriate external HV supply).  The
FemtoDAQ has been extensively tested with both SiPMs and PMTs.

\subsection{Triggering, Data Capture, and Recording}

\begin{figure}[!t]
\centering
\includegraphics[width=3.5in]{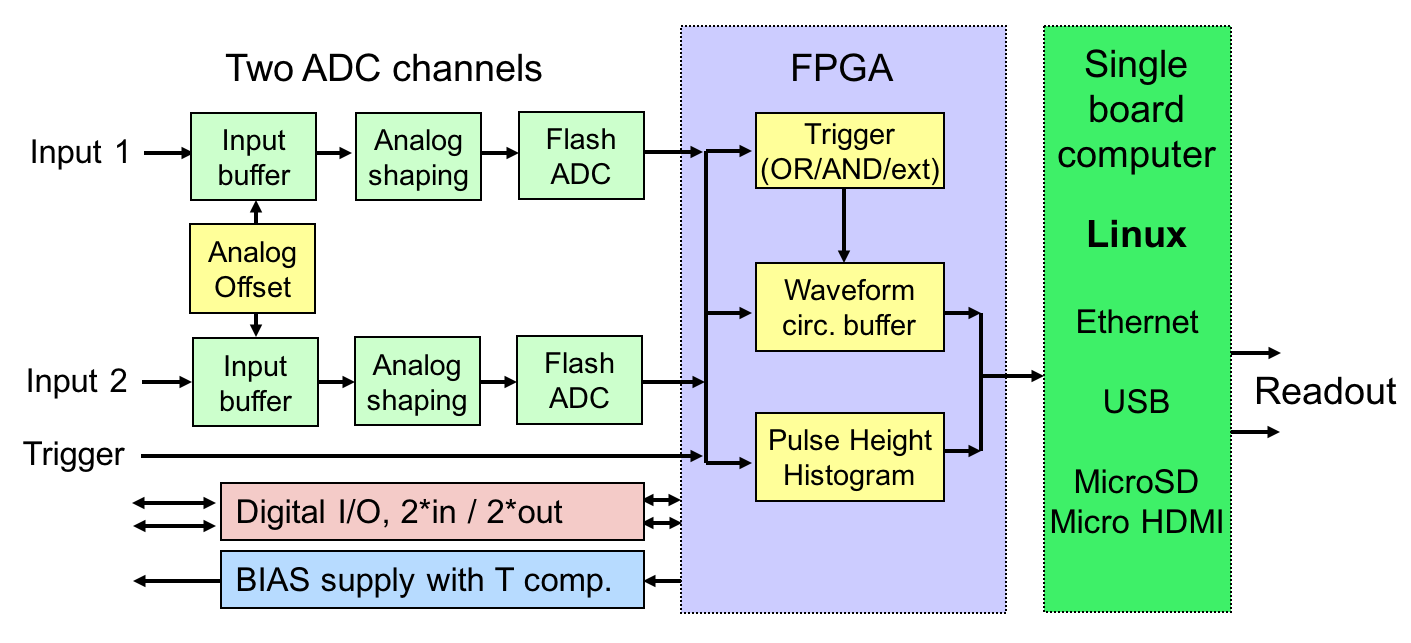}
\caption{Flow and control of sensor input to the FemtoDAQ.}
\label{fig:diagram}
\end{figure}

The flow and control of data in the FemtoDAQ is shown in
Fig.~\ref{fig:diagram}.  After digitization, a trigger condition is applied by
the onboard FPGA. The trigger thresholds are specified by the user and can be
defined independently for the two input channels or as a coincidence trigger on
both channels. One of the logic inputs also provides an optional external
trigger.

Captured waveforms are stored in a circular buffer with a storage length of
81.92~$\mu$s per waveform per channel.  Users can read out individual triggered
waveforms or instruct the FPGA to compute a pulse height histogram for all
triggers captured on each channel within a user-defined time period. Histograms
are stored with 12-bit resolution (4096 bins per channel). For histogramming
{\sl in situ}, the pulse height is analyzed by a rolling average with a
selectable bin size. Other DSP filters, such as baseline subtraction,
differentiation, etc. are also implemented in the FPGA.

The FemtoDAQ contains onboard flash storage with 4~GB capacity, shared with the
Linux OS installed on the BeagleBone Black computer. The use of solid-state
memory allows the device to be used in a variety of harsh environmental
conditions, including very high altitudes.

Users interact with the FemtoDAQ via the BeagleBone, which is powered
by a 1~GHz ARM processor with floating point hardware and ships with Debian
Linux running on board. The FemtoDAQ can be run as a standalone computer; an
HDMI connector is provided for connections to a local monitor, and a mouse and
keyboard can be connected via a USB type A connector.  A MicroSD socket is
available for storing acquired waveforms, histograms, and event files. In
addition, the network interface provided with the BeagleBone allows output to
be written to NFS-mounted remote disks.

Alternatively, the FemtoDAQ can be mounted as a peripheral device on a host
computer using available Ethernet or USB-2 connectors on the main board. Free
device drivers available from BeagleBone are used to mount the FemtoDAQ on
hosts running Windows, Linux, and Mac OS X; extensive tests have been performed
using Windows 8 and 10, Ubuntu 15.10, and Mac OS 10.11. 

\subsection{Programming and User Interface}

The FemtoDAQ presents the user with several programming interfaces which can be
customized for different applications. At the lowest level, users can control
hardware settings in the FPGA and input channels using commands sent to the
unit. These commands are used to adjust triggering and processing of waveforms
such as trigger levels, baseline subtraction, boxcar averaging, coincidence
conditions, etc. If the FemtoDAQ is powering one or two SiPMs, the command
interface can also be used to remotely adjust the bias voltage supplied by the
power board.

A Python module is provided to parse terminal input from the
user and pass commands to the onboard FPGA. The module is open-source and can
be customized or altered by the end user. This interface allows for a wide
variety of scripting and automation: for example, automatic timed measurements
with adjustments of bias voltage during the course of data taking.

\begin{figure}[!t]
\centering
\includegraphics[width=3.5in]{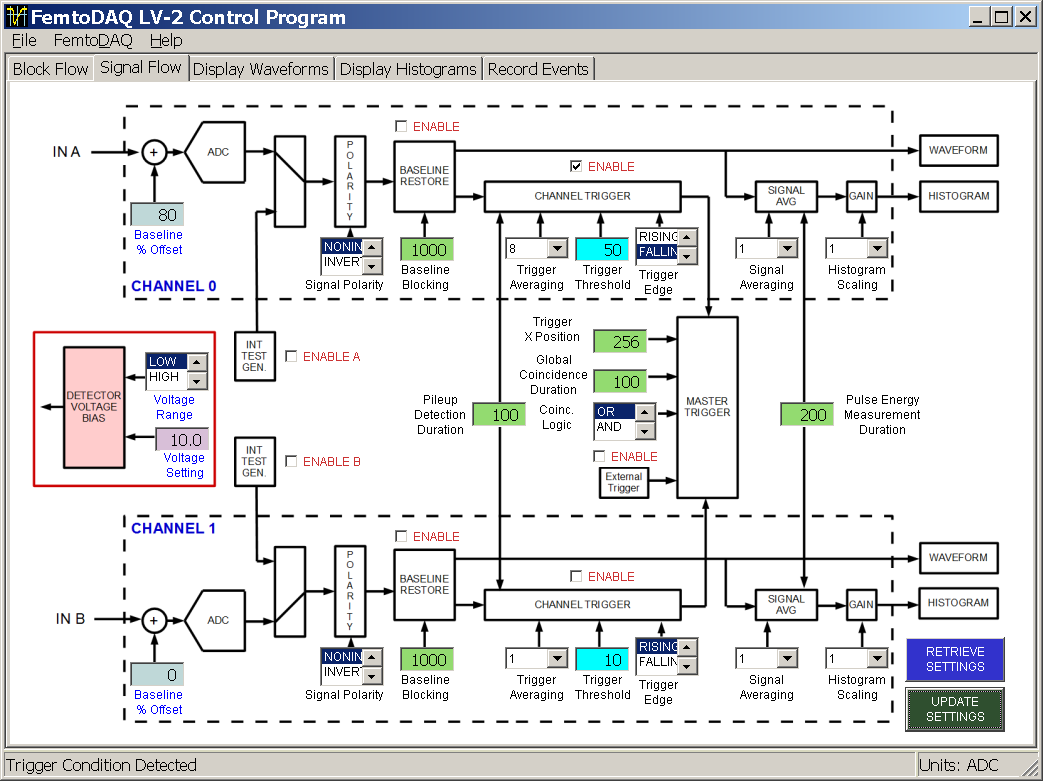}
\caption{Screenshot of the FemtoDAQ GUI control/flow window.}
\label{fig:gui}
\end{figure}

\begin{figure}[!t]
\centering
\includegraphics[width=3.5in]{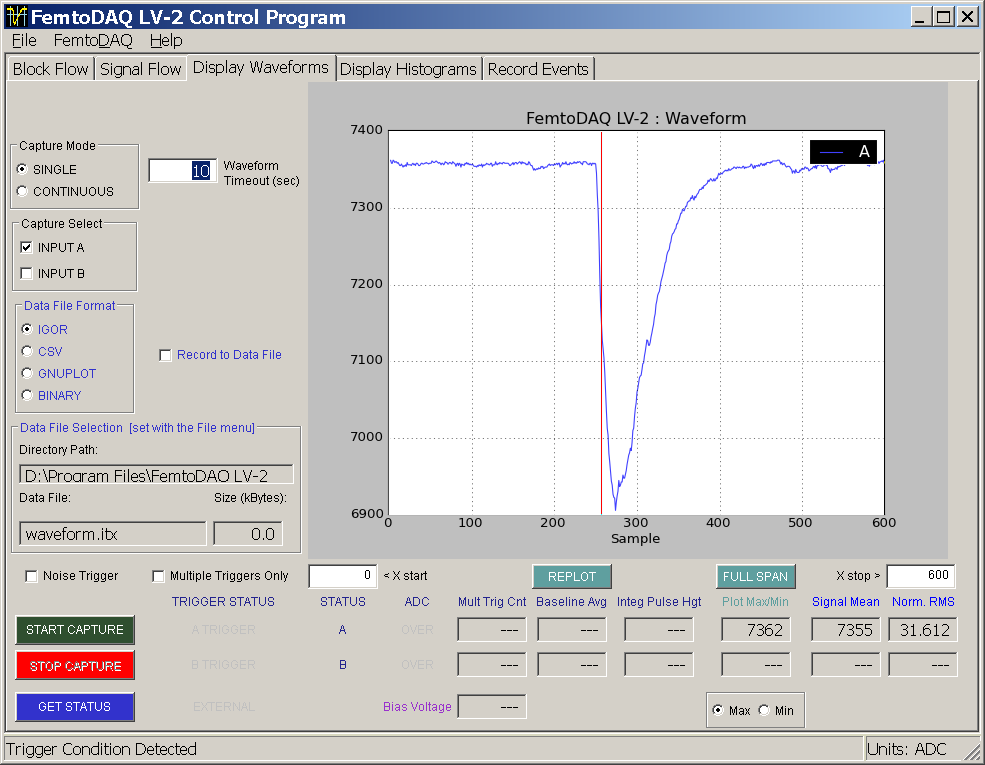}
\caption{Screenshot of the FemtoDAQ GUI showing a captured waveform.}
\label{fig:gui_waveform}
\end{figure}

While the FemtoDAQ is relatively easy to control via the command line, an
intuitive high-level graphical user interface (GUI) based on the wxPython
library is also provided.  The GUI, shown in Fig.~\ref{fig:gui}, allows users
to change the SiPM bias, enable internal test pulses, adjust baseline
correction and boxcar averaging, enable coincident triggers, and switch between
waveform capture mode and pulse height histogram mode. Online plotting
capabilities are also provided so that users can immediately view waveforms and
histograms (Fig.~\ref{fig:gui_waveform}).

For offline analysis, waveforms can be saved in one of several binary and text
formats -- e.g., comma-separated variable tables of ADC count versus clock --
and analyzed on the BeagleBone or saved to network disks or host computers.

\subsection{Example Application: Energy Resolution of LYSO}

To demonstrate the capabilities of the FemtoDAQ we provide data from two simple
studies: a nuclear/particle counting application in which a SiPM is coupled to
a LYSO scintillator, and a timing test of two SiPMs read out simultaneously in
both channels.

LYSO is a high-density scintillator commonly used in PET due to its high
efficiency for 511~keV photo-peak detection \cite{Du:2009}. It is also of
interest for National Security applications because of its high efficiency and
good timing properties \cite{Park:2013}. A challenge is its internal
radioactivity caused by traces of $^{176}$Lu.

\begin{figure}[!t]
\centering
\includegraphics[height=1.7in]{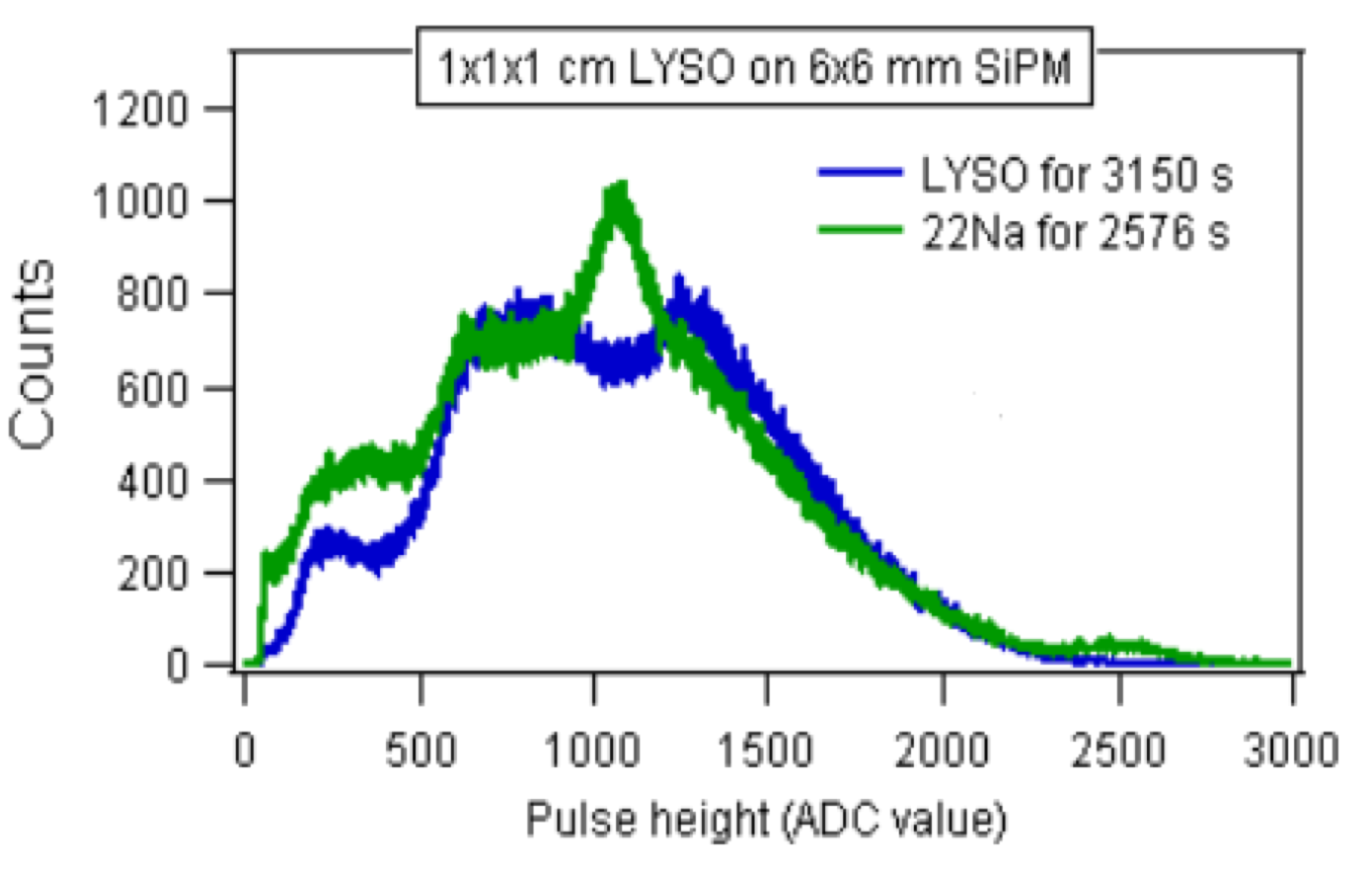}
\includegraphics[height=1.7in]{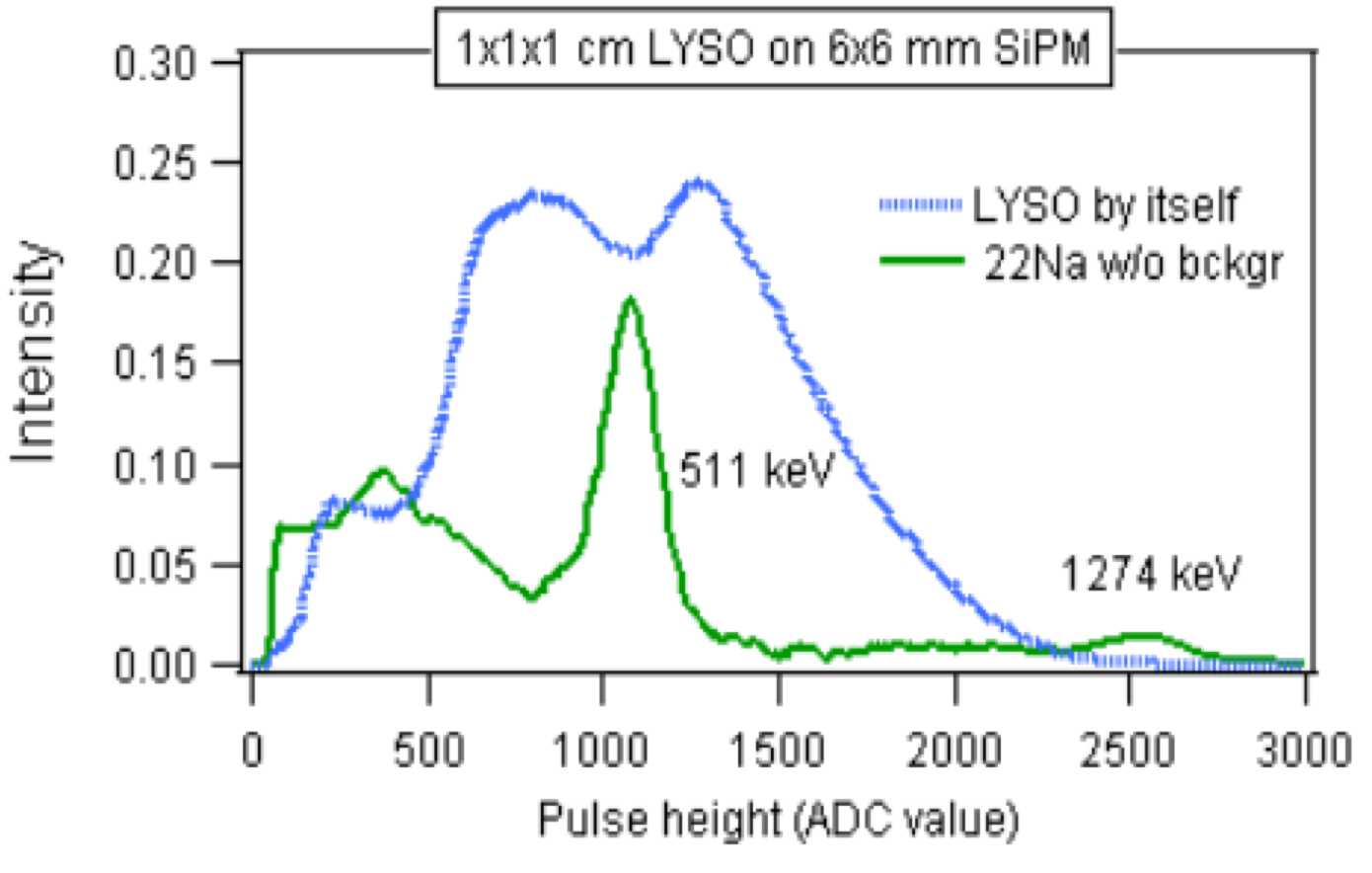}
\caption{Energy histogram of $^{22}$Na observed with 1~cm$^3$ LYSO (top) and
with background removed (bottom).}
\label{fig:lyso}
\end{figure}

Energy histograms from a 1~cm$^3$ LYSO scintillator are plotted in
Fig.~\ref{fig:lyso}. The scintillator was optically coupled to a single 6x6 mm
Series-C SiPM manufactured by SensL \cite{CSeries:2014}, and the output from
the SiPM was amplified by a factor of $10$ using a custom-made carrier board. A
$^{22}$Na $\gamma$ source was placed on top of the crystal, and the resulting
light signals observed by the SiPMs were used to self-trigger the FemtoDAQ and
produce a pulse height histogram. A ``free run'' pulse height histogram due to
the intrinsic $^{176}$Lu activity was then subtracted from the measurement of
the $^{22}$Na source with the same crystal. Both histograms were produced using
the {\sl in situ} histogrammming capabilities of the firmware.

The 511~keV peak observed in the scintillator was fit offline using a Gaussian
and determined to be 74.8~keV FWHM, or about 14.6\%. This peak resolution is
typical for LYSO \cite{Kimble:2002}.

\subsection{Example Application: SiPM Timing}

The second demonstration of the capabilities of the FemtoDAQ is a measurement
of the relative timing of two SiPMs observing a common light source.

\begin{figure}[!t]
\centering
\includegraphics[height=1.7in]{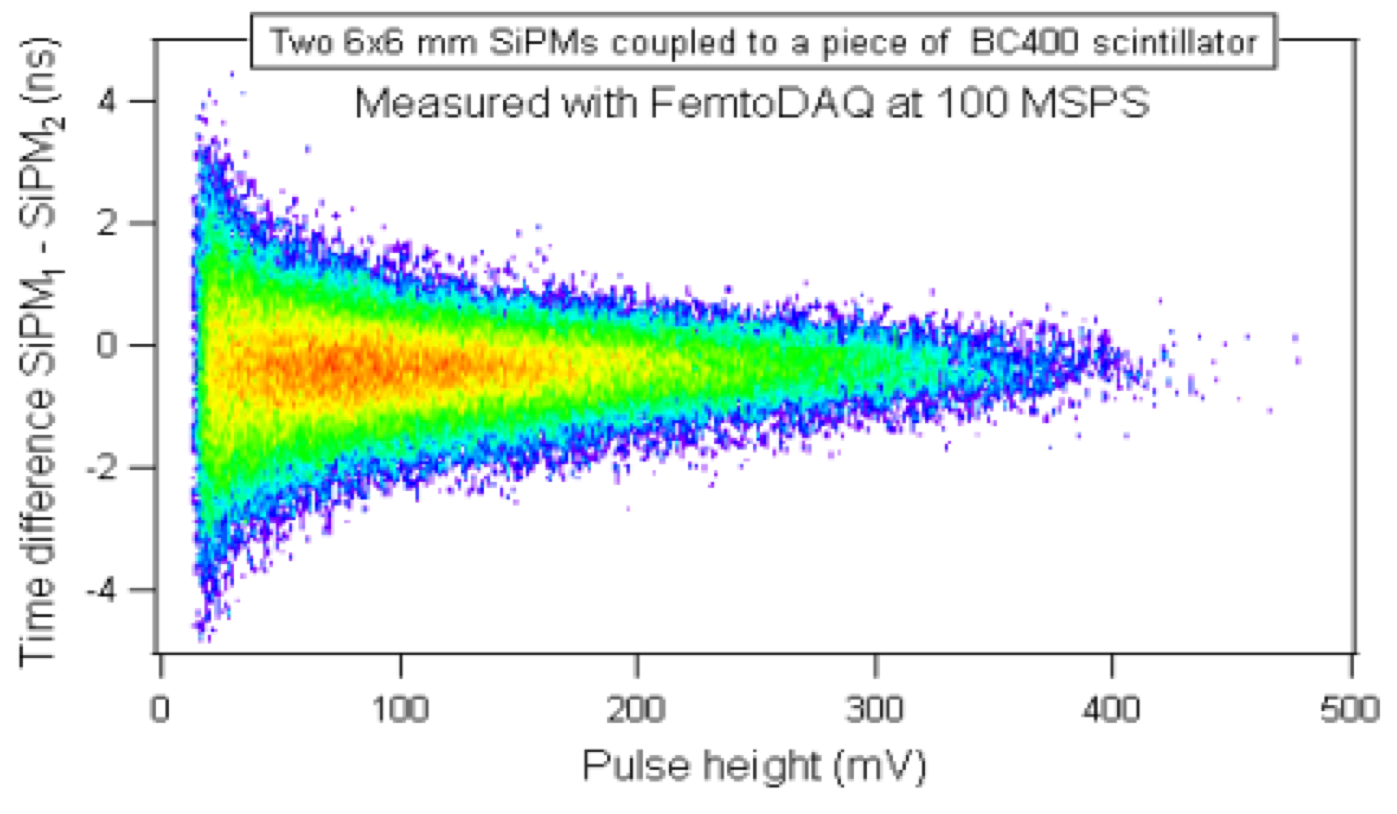}
\includegraphics[height=1.7in]{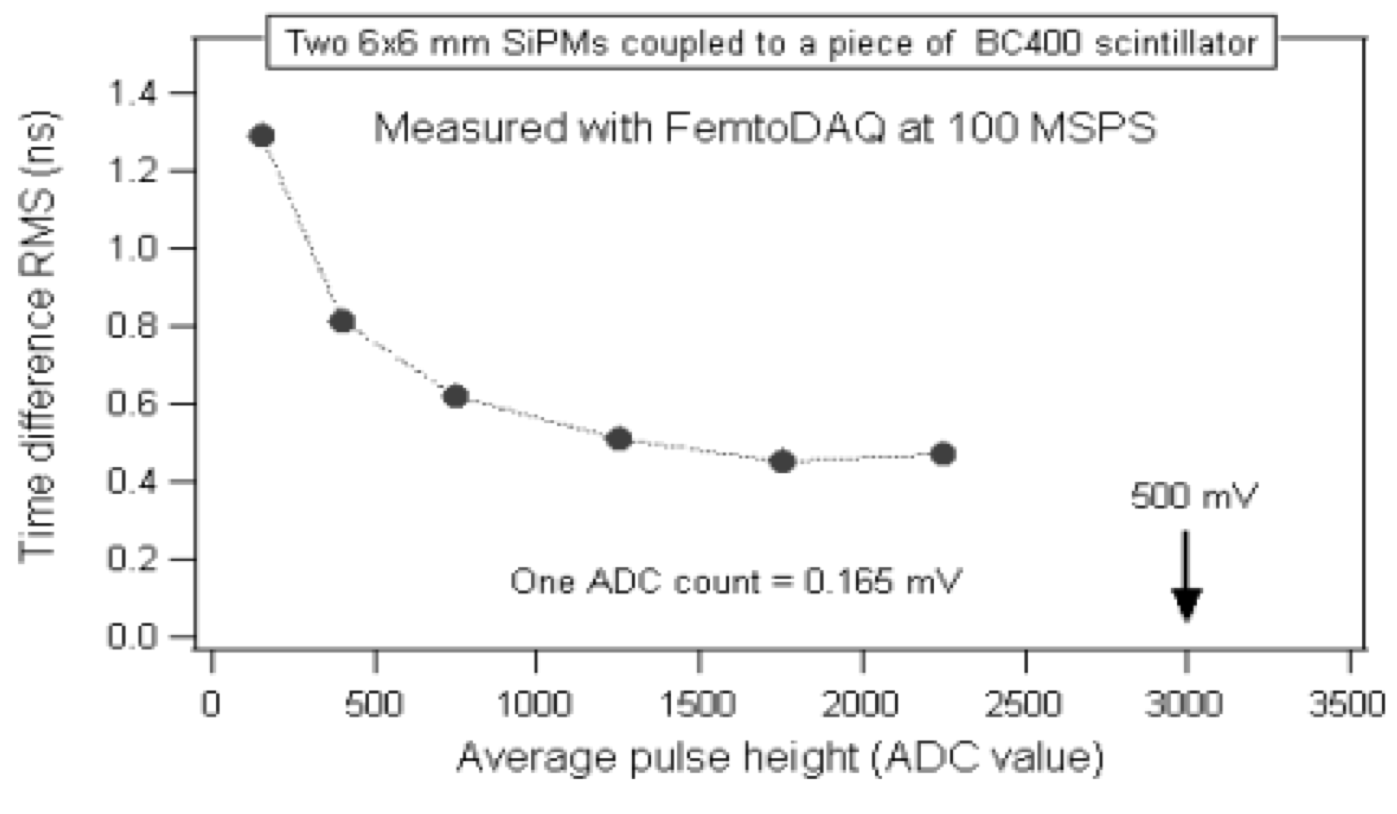}
\caption{Relative timing of two SiPMs observing the same $\alpha$ source (top),
showing improvement in the time difference as a function of pulse height
(bottom).}
\label{fig:timing}
\end{figure}

We measured the timing using digitized signals from two 6$\times$6~mm SiPMs
coupled to a single piece of BC400 irradiated with $\alpha$ particles from a
thick natural $^{232}$Th source. The 10$\times$ amplified signals were
digitized with two FemtoDAQ channels in self-triggered mode. A total of $10^5$
events were recorded, each consisting of two correlated waveforms from both
SiPMs. The event file was processed offline.

In Fig.~\ref{fig:timing} we show the time difference between SiPM$_1$ and
SiPM$_2$ plotted versus the average pulse height $(\text{PH}_1+\text{PH}_2)/2$.
The lower part of the figure shows the RMS projections of the 2-dimensional
distribution as a function of average pulse height. The largest pulse height
corresponds to $\alpha$ particles of about 6~MeV (in air) from $^{212}$Po decay
in the source.

We observed an improvement in the relative timing as a function of pulse
height; the highest amplitude events achieved 450 ps RMS, which is 4.5\% of the
sampling clock. The largest pulses correspond to about 15\% of the full 14-bit
ADC range of 16~384 ADC counts.

%\subsection{Pulse Discrimination with CsI}
%
%\begin{figure}[!t]
%\centering
%\includegraphics[height=1.7in]{pulse1_bw}
%\includegraphics[height=1.7in]{pulse2_bw}
%\caption{Particle discrimination using the FemtoDAQ.}
%\label{fig:pulse}
%\end{figure}
%
%A CsI(Tl) crystal was coupled to four 6$\times$6~mm SiPMs. The waveforms were
%recorded on disk and processed offline. Analyzing the fast and slow components
%and their ratios as a function of pulse height (Fig.~\ref{fig:pulse}) enables
%discrimination between $\alpha$ and $\gamma$ particles.

\section{The HAWC Gamma-Ray Observatory}\label{sec:hawc}

The HAWC Observatory is designed to observe the extensive air showers produced
when astrophysical $\gamma$ rays and cosmic rays interact in the atmosphere and
produce a high-energy particle cascade. HAWC is a collaboration of 30 academic
institutions and national labs in the US, Mexico, and Europe. The detector is
located 4100~m above sea level in Sierra Negra, Mexico, and is optimized to
detect $\gamma$ rays and cosmic rays between 100~GeV and 100~TeV.

The detector is a 20~000~m$^2$ array of 300 close-packed water Cherenkov
detectors (WCDs). Each WCD contains 200~kL of purified water.  When
relativistic charged particles from extensive air showers pass through the
water they produce ultraviolet Cherenkov light. Three 8'' hemispherical PMTs
and one 10'' PMT located at the bottom of each tank (for a total of 1200 PMTs
in the full array) are used to detect the Cherenkov photons. By combining the
timing information and spatial distribution of PMTs triggered by an air shower,
the arrival direction, energy, and type of the primary particle can be
identified.
In this manner, air showers produced by cosmic rays can be
filtered out of the data during offline processing, and the remaining $\gamma$
rays are used to produce sky maps of $\gamma$-ray sources.

\subsection{High-Energy Upgrade}

The construction of HAWC was completed in December 2014, and a high-energy
upgrade to the detector is now underway. To increase the sensitivity of the
observatory above 10~TeV, the collaboration will deploy 350 ``outrigger'' tanks
in a sparse array surrounding the 300 WCDs already at the site
\cite{Sandoval:2015}.

\begin{figure}[!t]
\centering
\includegraphics[width=3.5in]{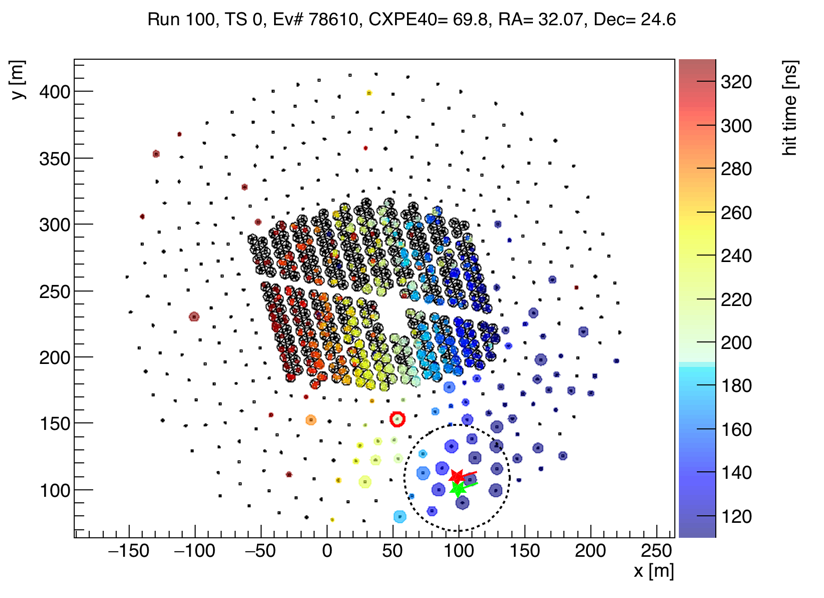}
\caption{Event display of a simulated $\gamma$-ray triggering the HAWC
detector. The observatory is viewed from above, with large black circles
showing the central 300 WCDs already deployed, and the small black
circles indicating the positions of the 350 outrigger stations currently under
development. Colored markers indicate PMTs triggered by the event. PMT timing
is indicated by the color scale, and the marker size is proportional to the
charge observed in each PMT.}
\label{fig:hawc_or}
\end{figure}

The purpose of the outriggers is to help localize high-energy air showers that
impact just outside the central array. Currently these air showers are poorly
reconstructed and must be removed from the analysis of cosmic-ray and
$\gamma$-ray events. However, with the outrigger tanks in place, the
``uncontained'' showers currently lost to quality cuts can be recovered. An
example of such an event is shown in Fig.~\ref{fig:hawc_or}.

The design of the outrigger stations and readout electronics is in progress
and deployment of the outrigger stations will begin in late 2016. The
outrigger upgrade is expected to increase the effective area of HAWC by a
factor of four above $10$~TeV.

\subsection{Testing SiPM Detectors for Use in HAWC}

The baseline design of an outrigger station calls for the use of an 8''
Hamamatsu R5912 hemispherical PMT, but the upgrade presents an excellent
opportunity to field-test new photosensor technologies such as SiPMs. SiPMs
have many advantages over PMTs as described in Section~\ref{sec:femtodaq}, and
are the focus of considerable manufacturer R\&D. The technology is experiencing
significant year-over-year improvements in quality and performance, and is an
excellent candidate to replace PMTs in water Cherenkov experiments in the near
future.

We have used the FemtoDAQ in the laboratory to characterize the basic
properties of SiPMs that are important for a photon counting experiment such as
HAWC -- for example, single photoelectron (PE) response, dark count rate, gain,
etc.  For this study we have focused on the SensL C-Series 60035-4P-EVB, a
$2\times2$ array of 3~mm silicon photosensors. The photon detection efficiency
of the sensor peaks at 40\% between 400 and 450~nm, and its gain ranges from
$10^6$ to $10^7$ as the overvoltage is increased from 1~V to 5~V.

\begin{figure}[!t]
\centering
\includegraphics[height=2in]{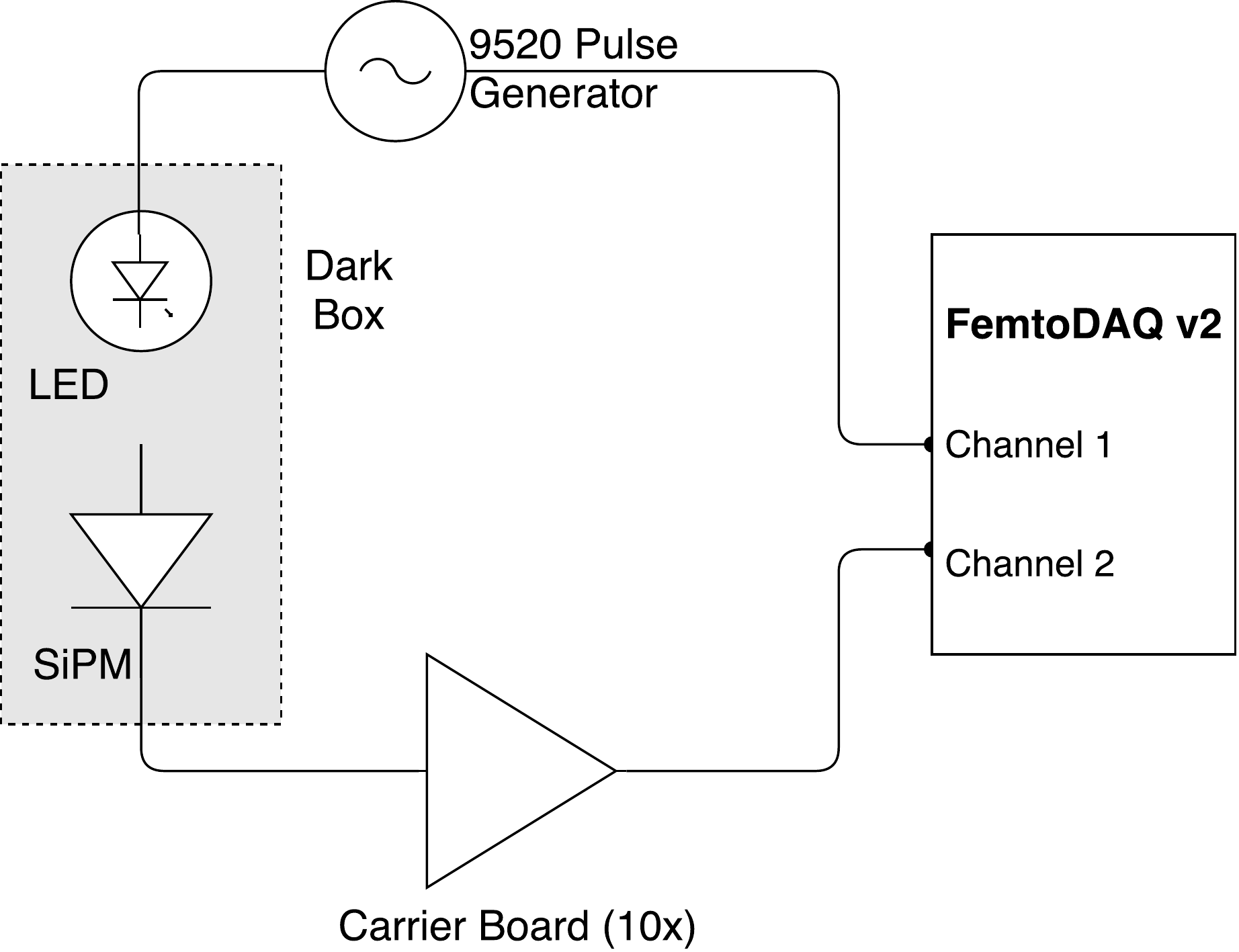}
\caption{Diagram of the setup used for single PE measurements of the SiPM
array.}
\label{fig:setup}
\end{figure}

Our setup for testing the photon-counting response of the SiPM array is shown
in Fig.~\ref{fig:setup}. The array is seated on a custom carrier board which
provides $10\times$ amplification. The board acts as a front end to the
FemtoDAQ, which provides a bias voltage via its internal power board.

The SiPM has been placed inside a dark box with a 5~mm UV (400~nm) LED. A
Quantum Composer 9520 pulse generator is used to pulse the LED with extremely
narrow ($\sim10$~ns) square pulses, so that on average only a few photons are
emitted per pulse. The generator output is split between the LED and Channel 1
of the FemtoDAQ, with the output from the SiPM sent to Channel 2.

To observe the single PE response of the SiPM, the FemtoDAQ has been set up to
trigger on coincident input from the split pulser signal (Channel 1) and the
SiPM (Channel 2). By gating the SiPM output in this manner we can effectively
eliminate dark counts from the triggered output. (Note that in the current
generation of SiPMs, the dark count rate can be considerable depending on the
overvoltage and temperature of the sensor. At 20~C, the dark count rate from
the C-Series array is typically of order 1~MHz.)

The SiPM output has a very long RC tail ($>1~\mu$s) because each cell in the
device is connected in series to a quenching resistor \cite{SiPM:2014}. The
tail can cause a pile-up effect when the photon rate (or dark count rate) is
too high, so we have chosen to remove it by numerically differentiating the
digitized output from Channel 2 of the FemtoDAQ. This is roughly analogous to
shaping the signal with a CR-RC circuit in a NIM crate.

\begin{figure}[!t]
\centering
\includegraphics[width=3.5in]{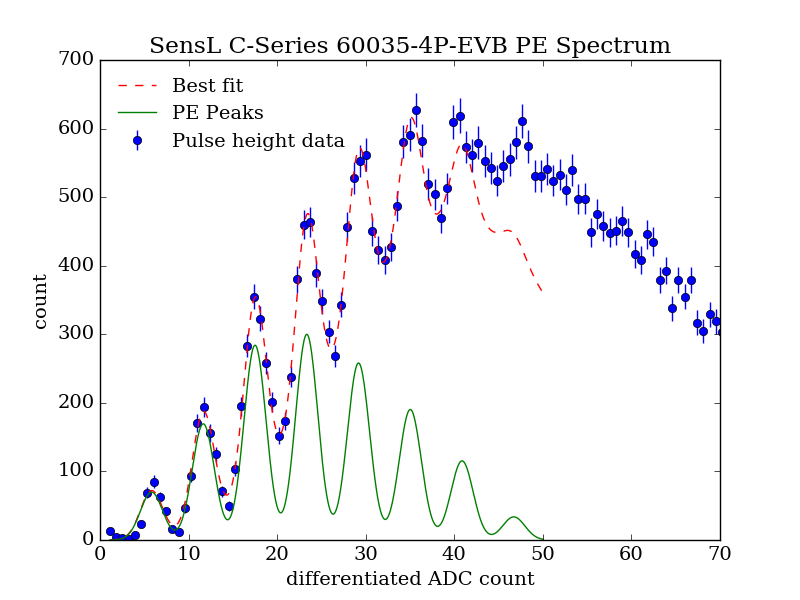}
\caption{Single and multiple PE pulse height spectrum of the SiPM
array (see text for details). The spectrum is fit to a set of Gaussian peaks
plus noise background (red curve); the green curve shows the Gaussian peaks
with the background removed.}
\label{fig:pe_data}
\end{figure}

After differentiation, the SiPM waveform is transformed from a set of rising
edges with exponential tails to a set of well-isolated peaks. The peaks have a
small undershoot which we compensate with offline processing (though this could
also be done internally by the FPGA). The pulse height histogram of the
processed waveform is plotted in Fig.~\ref{fig:pe_data}. Individual Gaussian
peaks from single and multiple PE events are clearly visible, while the dark
counts are strongly suppressed by the coincidence trigger. From the pulse
height histogram, we can easily set up calculations of the dark count rate and
relative gain as a function of bias voltage.

\section{Conclusion}\label{sec:conclusion}

The FemtoDAQ is a compact two-channel digitizer and signal processing unit
designed to read out a wide variety of commercial photosensors. Though it is
built atop the BeagleBone Black single-board computer, which is designed for
hobbyists, it is powerful enough to be employed in a laboratory setting. In
addition, its network capability and Python-based user interface allow for many
kinds of automated and/or remote measurements. We are currently using it to
characterize the performance of SiPMs for the outrigger upgrade of the HAWC
Observatory. 

The FemtoDAQ is also simple and robust enough for classroom use, and can be
operated easily by undergraduates and high school students. Educators with a
passing knowledge of Python can use the FemtoDAQ to design and carry out a
large number of classic university-level nuclear and particle physics
experiments, two of which were demonstrated in this paper. The compactness of
the device also suggests many interesting outreach activities,
%(balloon-borne measurements of cosmic rays immediately come to mind).  In addition to our
%laboratory use of the FemtoDAQ,
and we are working with local educators to design outreach programs.

% conference papers do not normally have an appendix

% use section* for acknowledgment
\section*{Acknowledgment}

WS acknowledges support in this work by the Department of Energy Office of
Science, Office of Nuclear Physics under grant numbers DE-SC0009543 and
DE-SC0013144. SB is supported by the Office of High Energy Physics under grant
number DE-SC0008475. We are indebted to Mr. David Hunter for implementing the
graphical user interface.

% trigger a \newpage just before the given reference
% number - used to balance the columns on the last page
% adjust value as needed - may need to be readjusted if
% the document is modified later
%\IEEEtriggeratref{8}
% The "triggered" command can be changed if desired:
%\IEEEtriggercmd{\enlargethispage{-5in}}

% references section

% can use a bibliography generated by BibTeX as a .bbl file
% BibTeX documentation can be easily obtained at:
% http://mirror.ctan.org/biblio/bibtex/contrib/doc/
% The IEEEtran BibTeX style support page is at:
% http://www.michaelshell.org/tex/ieeetran/bibtex/
%\bibliographystyle{IEEEtran}
% argument is your BibTeX string definitions and bibliography database(s)
%\bibliography{IEEEabrv,../bib/paper}
%
% <OR> manually copy in the resultant .bbl file
% set second argument of \begin to the number of references
% (used to reserve space for the reference number labels box)

% that's all folks
\end{document}